# The continuous net benefit: Assessing the clinical utility of prediction models when informing a continuum of decisions

Jose Benitez-Aurioles[1], Laure Wynants[2,3,4], Niels Peek[5], Patrick Goodley[6,7], Philip Crosbie[6,7], Matthew Sperrin[1]

[1] Centre for Health Informatics, University of Manchester, Manchester, United Kingdom

[2] Department of Epidemiology, Care and Public Health Research Institute, Maastricht University, Maastricht, The Netherlands

[3] Department of Development and Regeneration, KU Leuven, Leuven, Belgium

[4] Leuven Unit for Health Technology Assessment Research (LUHTAR), Department of Public Health and Primary Care, KU Leuven, Leuven, Belgium

[5] THIS Institute, University of Cambridge, Cambridge, United Kingdom

[6] Division of Immunology, Immunity to Infection & Respiratory Medicine, School of Biological Sciences, University of Manchester, Manchester, United Kingdom

[7] Manchester Thoracic Oncology Centre, Manchester University NHS Foundation Trust, Manchester, United Kingdom

## Abstract

Clinical prognostic models help inform decision-making by estimating a patient's risk of experiencing an outcome in the future. The net benefit is increasingly being used to assess the clinical utility of models. By calculating an appropriately weighted average of the true and false positives of a model, the net benefit assesses the value added by a binary decision policy obtained when thresholding a model. Although such 'treat or not' decisions are common, prognostic models are also often used to tailor and personalise the care of patients, which implicitly involves the consideration of multiple interventions at different risk thresholds.

We extend the net benefit to consider multiple decision thresholds simultaneously, by taking a weighted area under a rescaled version of the net benefit curve, deriving the continuous net benefit. In addition to the consideration of a continuum of interventions, we also show how the continuous net benefit can be used for populations with a range of optimal thresholds for a single treatment, due to individual variations in expected treatment benefit or harm, highlighting limitations of current proposed methods that calculate the area under the decision curve. We showcase the continuous net benefit through two examples of cardiovascular preventive care, comparing two modelling choices using the continuous net benefit.

The continuous net benefit informs researchers of the clinical utility of models during selection, development, and validation, and helps decision makers understand their usefulness, improving their viability towards implementation.

## Introduction

Clinical prediction models are used to determine a patient's diagnostic or prognostic risk. This can be useful when deciding whether patients should be screened, treated, monitored, or referred for diagnosis. Performance metrics such as the C-statistic or calibration plots can be used to assess the performance of these models[1,2] but are insufficient when determining if they will be useful in practice.

Decision curve analysis[3] is used to compare clinical utility of such models using the net benefit, a metric determined by an appropriately weighted sum of true and false positives. Models are evaluated by binarizing the risk scores at a clinically informed choice of threshold[4], with the decision curve plotting this threshold-specific net benefit for a range of thresholds. This method was originally designed for models informing a single clinical decision[3]. However, often a model's output is not used just to answer a single 'treat or not' question. For example, in cardiovascular disease prevention, a clinician might use QRISK[5] to decide whether to give a patient lifestyle recommendations, refer them to a smoking cessation programme, prescribe statins and/or antihypertensives, and increase monitoring frequency[6–8]. In such cases, a model is considered superior to another model if its corresponding decision curve is above the other model's curve over the entire range of relevant thresholds. Whilst you can interpret a difference in net benefit between two models for a particular threshold, you cannot meaningfully interpret the difference in net benefit of one model over many thresholds. This introduces difficulties in developing metrics that 'summarize' the net benefit over a range, and these approaches are subject to debate. However, a single-metric evaluation of net benefit across multiple thresholds could be useful when optimising or comparing models.

In this paper, we show why, in order to consider the previous formulation of the area under the net benefit curve valid, an assumption of equal value of true positives across the population is needed. We introduce a new metric, called the continuous net benefit, derived from the original conditional utility formula, and discuss its use, connecting it to performance metrics like the likelihood function and Brier score. The continuous net benefit requires an appropriately chosen weighting function across thresholds to be specified. We discuss the potential use of continuous net benefit to inform developers of clinical utility during model selection and validation, including examples of its application.

## Introduction to the net benefit

We explore this problem through an example in cardiovascular prognosis. Consider a population with a binary outcome of interest $Y$ which indicates whether an individual will have a cardiovascular event like a heart attack in the next ten years. A model is developed so that the risk of the outcome is estimated for each individual. Individuals with model scores above a threshold probability $t$ are prescribed statins, and the rest are not. The performance of that policy is summarised by its per capita true positives $TP(t)$ (number of people given statins who would have had a cardiovascular event if they had not been prescribed statins, divided by population size), false positives $FP(t)$ (people treated which would not have had the event if they had not been prescribed statins), false negatives $FN(t)$ (people who were not treated and who would then have the event) and true negatives $TN(t)$ (people who were not treated and would then not have the event). If the utilities $a$, $b$, $c$, and $d$ represent the benefit of each of these elements, the utility, conditional on the population, of the model is defined as:

$$U(t) = aTP(t) + bFP(t) + cFN(t) + dTN(t) \qquad (1)$$

The net benefit $NB$ is a simplified version of the utility which does not require the estimation or assumption of the four utilities $a$, $b$, $c$ and $d$ in order to be calculated. It is a linear transformation of $U$, so that if one model has a higher $NB$ than another, it also has a higher $U$. To derive the net benefit, the best choice of threshold, or clinically optimal threshold $t^*$ is determined by $\frac{1-t^*}{t^*} = \frac{a-c}{d-b}$, so that knowing the value of the utilities determines $t^*$, and thus knowing $t^*$ is informative of the utility values[3]. Rescaling the benefit of giving a patient statins when they would have had a cardiovascular event (i.e., $a - c$) to be 1, the net benefit becomes:

$$NB(t^*) = TP(t^*) - \frac{t^*}{1-t^*} FP(t^*) \qquad (2)$$

The optimal threshold $t^*$ can be determined from initial choices of the utilities $a$ to $d$, or more commonly through clinical knowledge and practice. If the optimal threshold $t^*$ is not known or varies across the population, the net benefit can be plotted across a reasonable range of values for $t^*$. This plot is known as the decision curve.

## Limitations of previous area under the net benefit curve metrics

A previous motivation for calculating the area under the net benefit curve has been the idea that individuals, due to different underlying utilities ($a^*$, $b^*$, $c^*$ and $d^*$), have different optimal thresholds, so that models need to be beneficial across a range of thresholds. For example, there will always be differences between individual patients in the extent to which they benefit from treatment and get harmed by side-effects. One approach to account for this distribution of optimal thresholds $p(t^*)$ is to check the decision curve throughout a previously determined range of reasonable thresholds (i.e., the range of $t^*$ for which we think $p(t^*)$ is non-negligible). If the net benefit of a model is better than another policy across the entire range, the model is considered more beneficial[4]. This, and all decision curve-based approaches, assumes that, across the population, the threshold is a random variable $T^*$ independent of the performance of the model at that threshold, so that the overall estimate of the performance in terms of $TP(t^*)$ and $FP(t^*)$ is not biased compared to the subgroup-specific performance in the individuals for which the threshold is $T^* = t^*$.

The area under the net benefit curve[9,10] has been proposed as a measure of the clinical benefit of a model in a population with a distribution of optimal thresholds $T^*$ by calculating its expected net benefit. It uses an estimated (or assumed) distribution of the optimal threshold $p(t^*)$ to calculate the expected net benefit in a population:

$$AUNB = \int_0^1 p(t^*)\left(TP(t^*) - \frac{t^*}{1-t^*}FP(t^*)\right)dt^* \qquad (3)$$

Here we highlight an underlying assumption of this equation which, to our knowledge, has not been previously discussed in the literature. Because the net benefit is in units of true positives, as mentioned in the previous section, so $a^* - c^* = 1$, the expression (3) is only an expectation of the utility when assuming that true positives are equally beneficial for all patients. The difference in optimal thresholds across the population is thus entirely dictated by variations in the false positive harm $(b^* - d^*)$ in each individual. This is unlikely to be true, as expected efficacy of treatment may drive differences in personal threshold. For example, statins are less effective in some groups[11], and these patients will have a lower true positive benefit $a^* - c^*$.

While this is not an issue when examining the decision curve, as it is meant to be compared at every threshold (it is read 'vertically'), it becomes an issue when the net benefit is integrated across a range (read 'horizontally'). Let's for example consider an alternative assumption, where the harm of a false positive $b^* - d^*$ is considered constant and equal to 1, so that the net benefit equation becomes $NB_{alt} = \frac{1-t^*}{t^*}TP(t^*) - FP(t^*)$, and the area under the net benefit is:

$$AUNB_{alt} = \int_0^1 p(t^*)\left(\frac{1-t^*}{t^*}TP(t^*) - FP(t^*)\right)dt^* \qquad (4)$$

For any single value of threshold $t^*$, $NB$ and $NB_{alt}$ are equivalent, in that if a model has higher net benefit than another in $NB$, it also has higher net benefit in $NB_{alt}$. This is however not true in terms of the integral as one model might have a higher $AUNB$ and lower $AUNB_{alt}$ than another.

In general, knowing the distribution of optimal thresholds $p(t^*)$ in a population of interest is not sufficient to calculate an expected or total benefit of a model in this population without additional knowledge or assumptions. To understand why this is the case, imagine a population of two groups $G_1$ and $G_2$ where each makes up half of the population. A model is used to inform a particular diagnostic decision in both groups, with equal calibration and discrimination in $G_1$ and $G_2$. Consider that in group $G_1$, there is no real value in diagnosis, nor harm in over-diagnosis, so that $a_{G_1}^* - c_{G_1}^*$ and $b_{G_1}^* - d_{G_1}^*$ are very small, but balanced so that $t_{G_1}^* = \frac{d_{G_1}^* - b_{G_1}^*}{d_{G_1}^* - b_{G_1}^* + a_{G_1}^* - c_{G_1}^*} = 10\%$. On the other hand, the decision is 'life-or-death' for group $G_2$, with large benefit in diagnosing the condition and serious harm in incorrectly flagging someone as positive, so $a_{G_2}^* - c_{G_2}^*$ and $b_{G_2}^* - d_{G_2}^*$ are very large, and balanced so that $t_{G_2}^* = 11\%$. A useful measure of the benefit of the model is not one that averages over the benefit of the model at 10% and 11% with equal weighting. In reality, decisions in group $G_1$ are inconsequential, and decisions in group $G_2$ are very important. It is thus important not only to know the distribution $p(t^*)$ across the population, but also to have more information of the relative scale of the utilities $a^* - c^*$ and $b^* - d^*$, in order to know how useful a model is for a population with varying thresholds.

## Combining the net benefit of two decisions

Before approaching the problem of single treatment with distributions of optimal thresholds, we focus on the related issue of combining the net benefit of multiple decisions. For example, we would now like to assess the net benefit of a cardiovascular model when both informing the prescription of statins and the enrolment of patients into a lifestyle intervention programme. It is often not necessary to combine the benefit of two interventions, as their benefit can be easily considered separately, and, in fact, different models can be used for each decision[12]. However, considering how to do this is a useful intermediary step for latter reasoning.

We assume that both decisions to intervene depend on the risk of the same outcome. The two interventions have different optimal thresholds, $t_1^*$ and $t_2^*$, different utilities $a_1$, $b_1$, $c_1$, $d_1$ and $a_2$, $b_2$, $c_2$ and $d_2$, different corresponding conditional utilities $U_1$ and $U_2$, and net benefit values of $NB_1 = NB(t_1^*)$ and $NB_2 = NB(t_2^*)$. We assume that the effects of the two interventions do not interact, or that at least the utilities of the second intervention are in terms of the added benefit and harm with respect to the first, so that the conditional utility of both model-informed decisions is:

$$U_{1+2} = U_1 + U_2$$
$$= a_1 TP(t_1^*) + b_1 FP(t_1^*) + c_1 FN(t_1^*) + d_1 TN(t_1^*)$$
$$+ a_2 TP(t_2^*) + b_2 FP(t_2^*) + c_2 FN(t_2^*) + d_2 TN(t_2^*) \qquad (5)$$

The goal is to find an expression of the net benefit $NB_{1+2}$ (i.e., the benefit of a model across the two interventions) so that, if and only if a model has a higher $U_{1+2}$ than another in a population, it also has a higher $NB_{1+2}$.

The first intuition for calculating the total net benefit $NB_{1+2}$ is to choose $NB_{1+2} = NB_1 + NB_2$. However, this is problematic, as $NB_1$ has as unit the benefit of treating a true positive with statins (so $a_1 - c_1 = 1$), and $NB_2$ has as unit the benefit of enrolling a true positive in a lifestyle intervention programme ($a_2 - c_2 = 1$). The treatment effect of these two interventions is not equal ($a_1 - c_1 \neq a_2 - c_2$), and thus, the sum of both net benefits cannot be interpreted as the total net benefit of the model. Instead, the ratio between the value of both interventions, $\frac{a_2 - c_2}{a_1 - c_1}$, needs to be estimated in order to calculate the total net benefit as:

$$NB_{1+2} = NB_1 + \frac{a_2 - c_2}{a_1 - c_1} NB_2 \qquad (6)$$

In the cardiovascular example, if we estimate that enrolment in a lifestyle intervention programme is half as effective in reducing cardiovascular events compared to treating a patient with statins, the total net benefit would be $NB_{1+2} = NB_1 + \frac{1}{2} NB_2$. Here, the unit of $NB_{1+2}$ is the benefit of treating a positive patient with statins. Any expression of $NB_{1+2}$ up to a rescaling and offset factor can be used, as this doesn't change the order of evaluated models.

## The continuous net benefit

In reality, cardiovascular prognostic models like QRISK are used for a very wide range of potential treatment strategies and monitoring plans that are generally impossible to enumerate in advance. We model this situation as a continuum of interventions, each with a particular optimal threshold $t'$, so that, at each threshold $t'$, the corresponding treatments carry an added true positive utility of $a'$ and false positive utility of $b'$. The total utility across all potential treatments for the patient is thus the integral of the conditional utility function:

$$U_{cont} = \int_0^1 a'TP(t') + b'FP(t') + cFN(t') + dTN(t') \, dt' \qquad (7)$$

Where $c$ and $d$ are constants, as they relate to the absence of any treatment. Writing $FN(t') = \pi - TP(t')$ and $TN(t') = 1 - \pi - FP(t')$, where $\pi = P(Y = 1)$ is the incidence of the outcome in the population, we get:

$$U_{cont} = d + (c-d)\pi + \int_0^1 (a'-c)TP(t') + (b'-d)FP(t')dt' \qquad (8)$$

The term $d + (c-d)\pi$ is equal for all policies or models, and can thus be ignored when comparing models. For any threshold $t'$, we still have $\frac{1-t'}{t'} = \frac{a'-c}{d-b'}$, and thus:

$$NB_{cont} = \int_0^1 \omega_{cont}(t') (\frac{TP(t')}{t'} - \frac{FP(t')}{1-t'}) \, dt' \qquad (9)$$

The sum $\frac{TP(t')}{t'} - \frac{FP(t')}{1-t'}$ is equal, up to a constant, to the sum of the net benefit in the treated and that in the untreated[13] for an intervention with optimal threshold $t'$. The weighting function $\omega_{cont}(t') = \frac{1}{\frac{1}{a'-c} + \frac{1}{d-b'}}$ is the harmonic mean between the utility of identifying a true positive $a' - c$, and avoiding a false positive $d - b'$. We call this weighting function the 'importance' of a threshold, as it, in practice, determines how much the benefit (and thus performance) of a model at a particular threshold should be weighed against the benefit on other thresholds. If a normalisation constant is chosen so that $\int_0^1 \frac{\omega_{cont}(t')}{t'} d\, t' = 1$, the unit of the continuous net benefit is still true positives (though it will be combined true positives, i.e., the benefit that one patient with $Y = 1$ gets when managed as a high-risk patient for all decisions).

Generally, this weighting function is high only when both the utilities $a' - c$ and $d - b'$ are high, and is low if either of the two is low. For example, consider possible treatments for cardiovascular disease, and the quality-adjusted life years (QALYs) added or lost due to true and false positives. We expect the benefit of true positives to range between 5 and 15 QALYs compared to not intervening, and the harms of false positives to be lower, between 0.1 and 1 QALYs lost compared to not intervening, as commonly used preventive interventions in cardiovascular care do not usually carry serious side-effects. As seen in Figure 1, in this case, the weighting is mostly determined by the severity of the harm of false positives, as it is the smaller effect in the harmonic mean, meaning that interventions with worse potential false positive harm should be weighted more than those with relatively low harm.

Some metrics commonly used to assess the performance of prediction models can be related to the continuous net benefit through specific choices of weighting function. For example, consider the choice of a uniform function $\omega_{cont}(t') = 1$. The chosen 'importance' of all thresholds is thus equal. In this case, $NB_{cont} = \int_0^1 \frac{TP(t')}{t'} - \frac{FP(t')}{1-t'} dt'$, and the difference between the continuous net benefit of two models $M_1$ and $M_2$ is equivalent to the difference between their likelihoods, with:

$$\omega_{cont}(t') = 1 \rightarrow NB_{cont}(M_1) - NB_{cont}(M_2) = \mathcal{L}(M_1) - \mathcal{L}(M_2) \qquad (10)$$

Here, $\mathcal{L}$ is the likelihood function of the models given the observed data. The derivation of this result is shown in Supplementary 1.1. Models are often fitted to maximise the likelihood function given the data, which could intuitively be interpreted as choosing the best model when considering each threshold equally important. On the other hand, if a weighting function is chosen to be a symmetric parabola $\omega_{cont}(t') = t'(1 - t')$ so that $NB_{cont} = \int_0^1 t'(1 -$ $t')(\frac{TP(t')}{t'} - \frac{FP(t')}{1-t'})dt'$, the difference between the continuous net benefit of two models is equivalent to the negative Brier score difference between them, with:

$$\omega_{cont}(t') = t'(1-t') \rightarrow NB_{cont}(M_1) - NB_{cont}(M_2)$$

$$= \frac{1}{2}\big(-Brier(M_1) + Brier(M_2)\big) \qquad (11)$$

The derivation of this result is shown in Supplementary 1.2. In practice, it is uncommon for thresholds to lie close to 0.5, with most clinical decisions having optimal thresholds closer to 0 or 1, explaining why the Brier score has been found to be a suboptimal metric to represent clinical benefit[14]. Finally, the binary net benefit of a model can be recovered by using a point mass as the weighting function $\omega_{cont}(t') = \delta(t' - t^*)$, giving $NB_{cont} = \int_0^1 \delta(t' - t^*) \big(\frac{TP(t')}{t'} - \frac{FP(t')}{1-t'}\big) dt' = \frac{TP(t^*)}{t^*} - \frac{FP(t^*)}{1-t^*}$.

## Considering single interventions with varying thresholds using the continuous net benefit

We revisit the motivation of previous literature of calculating an expected net benefit in a population with a distribution of utilities $a^*$, $b^*$, $c^*$ and $d^*$, and a corresponding distribution of optimal threshold $\frac{1-t^*}{t^*} = \frac{a^*-c^*}{d^*-b^*}$. In that case, the expected utility of the population is:

$$U_E = \int_0^1 p(t^*)\big(a^*TP(t^*) + b^*FP(t^*) + c^*FN(t^*) + d^*TN(t^*)\big) dt^* \qquad (12)$$

Following a similar derivation to that from equations (7) to (9), detailed in Supplementary 1.3, the expected net benefit across the population can be obtained as:

$$NB_E = \int_0^1 \omega_E(t^*)(\frac{TP(t^*)}{t^*} - \frac{FP(t^*)}{1-t^*}) dt^* \qquad (13)$$

Where the weighting function $\omega_E(t^*) = \frac{p(t^*)}{\frac{1}{a^*-c^*}+\frac{1}{b^*-d^*}}$ includes both the distribution of the thresholds across the population and the utilities for each of these thresholds. If the weighting function is normalised so that $\int_0^1 \frac{\omega_E(t^*)}{t^*} d\,t^* = 1$, the unit of the net benefit will be the averaged true positive across everyone in the population.

## Example 1: Developing a cardiovascular risk prognostic model for multiple interventions

We showcase the use of the continuous net benefit through an example. We develop and validate two logistic regression models to predict the risk than an individual will have a cardiovascular disease event in the next 10 years, using data from the Framingham Heart Study[15]. The first model only includes variables that can be quickly provided by the patient (age, sex, smoking status and prevalent cardiovascular disease), while the second model uses a more extensive set of predictors.

We evaluate the model for two separate sets of decisions: 1) the prescription of statins, evaluated at the optimal threshold of 10% used in UK practice[6], and 2) overall patient management and lifestyle recommendations. The first is evaluated by placing a point mass at the 10% threshold, while the second is evaluated using the continuous net benefit, choosing as weighting function $\omega_{cont}$ a half-Gaussian, corresponding to a full-Gaussian with mean of 10% and standard deviation of 2% with only non-zero values below 10%. This weighting function was chosen after discussion with a clinician, after which we determined that usually clinicians will attempt to reduce the risk of a patient through lifestyle interventions and more frequent monitoring before their risk reaches 10%. We believe that these interventions are especially 'aggressive' as the patient gets closer to the threshold of 10%, and carry higher risks, like those of exercise-related injuries or disengagement from healthcare. In line with how the weighting function changes due to variation in the false positive harm in this context, as shown in Figure 1, we chose an increasing function up to the threshold of 10%, at which clinicians start prescribing statins and stop modifying their care. Details on the development and validation of the model are included in Supplementary 2. Full code, along with R code showcasing how to calculate the continuous net benefit, can be accessed through GitHub (https://github.com/jaurioles/continuousNetBenefit). Confidence intervals (95%CI) and optimism corrections for the net benefit and continuous net benefit are calculated through bootstrapping[17].

The decision curves of the models, as well as the weighting function chosen, are plotted in Figure 2a, and the net benefit of both sets of interventions is shown in Figure 2b. For overall patient management (excluding statins), the continuous net benefit of the model with fewer predictors, in combined true positives per 100 people, is 7.7 (95%CI: 6.8 – 8.5) while for the model with more predictors the continuous net benefit is 8.2 (7.4 – 9.1), with a difference between the two of 0.5 (0.3 – 0.8). Generally, the continuous net benefit was in agreement with the decision curve, but quantified the difference in net benefit of the two models across the range, which can be helpful for further model optimisation or for overall comparisons of the two models.

Example 2: Developing a cardiovascular risk prognostic model for one intervention with a distribution of thresholds

In this second example, we seek to validate the same models in the same data, but instead consider that not all individuals will be prescribed statins when they are over the risk of 10%. Usually, the range of thresholds at which cardiovascular models are evaluated in decision curve analysis goes between 0% and 20%[5], although higher thresholds might be considered, due to the relative inefficacy of statins in some populations[11]. We model the distribution of optimal thresholds $p(t^*)$ to be a log normal distribution with mean of 10% and standard deviation of 3%, plotted in Figure 3a. Given that we think that the main variation of thresholds is due to changes in the relative risk reduction achieved by statins in different people, we assume that the harms due to false positives are constant ($d - b = 1$), so that we use equation (4) to calculate an expected net benefit across the population. If we wanted to consider variation in false positive harm in our analysis (as in reality, some patients value more negatively false positive harm of preventative treatment), equation (13), with an appropriate weighting function, would need to be used.

The continuous net benefit, shown in Figure 3b, in units of average true positives per 100 people, is 3.9 (2.9 – 4.9) when all patients are treated, 7.1 (6.3 – 8.0) for the model using less predictors, and 7.8 (7.0 – 8.7) for the model using all predictors. With our assumptions, we record a difference in the continuous net benefit between the two models of 0.7 (0.5 – 0.9). Assuming all individuals to have an optimal threshold of 10% gives the same net benefit difference, with wider confidence intervals, of 0.7 (0.3 – 1.0). Although, in this case, the two assumptions lead to the same difference in net benefit between the two models, the values in each model and the confidence interval of the difference vary.

Discussion

In this paper, we present an approach to assess the clinical utility of prognostic models in situations where the model is used for overall patient management, or situations where we believe that a single intervention's optimal threshold varies across the population. Summarising the total net benefit as a single measure showed the overall usefulness of prediction models in a cardiovascular problem.

A potential criticism of attempting to summarise the decision curve as a single metric is that they average over potential meaningful differences in the ranking of models across the decision curve. When informing two decisions, or the same decision for subgroups of a population, different models could be superior for the respective thresholds, and one model should not be chosen for both. We instead suggest using the continuous net benefit in situations where there is no clear separation between well-defined decisions, or for which clinicians couldn't in practice identify patients from the different subgroups at the time of model deployment (as is the case when subgroups aren't well defined, clinicians implicitly understanding the risk and benefit profile of each patient). The continuous net benefit should be used alongside decision curve analysis, not instead of it, similarly to how calibration-in-the-large and the calibration slope are reported alongside calibration plots.

An increasing number of clinical prediction models are being created, with a very limited number being implemented in practice[18,19]. In part, this is due to the large number of available models, and to the challenges faced by policymakers when inferring whether models would be clinically useful and beneficial to patients. Our work can be useful for decision makers when considering the use of models beyond a single systematic clinical decision, enabling them to assess the overall clinical value of an algorithm. The continuous net benefit could also be useful for researchers when comparing modelling choices, when needing a metric to optimise, or when summarising the decision curve of a model over multiple validation studies using meta-analysis[20]. Using reasonable assumptions or estimates for the weighting function can be a useful first step before carrying out more involved cost-benefit analysis. Further work could investigate how assumptions around the independence of optimal thresholds and predictors can be relaxed with subgroup analysis, apply the continuous net benefit to real model development examples, or explore different approaches to choose the weighting function.

## Acknowledgement

JBA is the receipt of the studentship awards from the Health Data Research UK-The Alan Turing Institute Wellcome PhD Programme in Health Data Science (Grant Ref: 218529/Z/19/Z). We acknowledge support of the UKRI AI programme, and the Engineering and Physical Sciences Research Council, for CHAI - Causality in Healthcare AI Hub [grant number EP/Y028856/1]. The research was carried out at the National Institute for Health and Care Research (NIHR) Manchester Biomedical Research Centre (BRC) (NIHR203308).

We thank Brian McMillan for providing helpful comments during the conception of this work.

## Data and computing code availability

Data used in this study is publicly available, as part of the 'riskCommunicator' R repository. Full code for this work has been uploaded to GitHub (https://github.com/jaurioles/continuousNetBenefit).

## Bibliography

1. Steyerberg EW, Vergouwe Y. Towards better clinical prediction models: seven steps for development and an ABCD for validation. *European Heart Journal*. 2014;35(29):1925-1931. doi:10.1093/eurheartj/ehu207

2. Steyerberg EW, Vickers AJ, Cook NR, et al. Assessing the performance of prediction models: a framework for traditional and novel measures. *Epidemiology*. 2010;21(1):128-138. doi:10.1097/EDE.0b013e3181c30fb2

3. Vickers AJ, Elkin EB. Decision Curve Analysis: A Novel Method for Evaluating Prediction Models. *Med Decis Making*. 2006;26(6):565-574. doi:10.1177/0272989X06295361

4. Van Calster B, Wynants L, Verbeek JFM, et al. Reporting and Interpreting Decision Curve Analysis: A Guide for Investigators. *Eur Urol*. 2018;74(6):796-804. doi:10.1016/j.eururo.2018.08.038

5. Hippisley-Cox J, Coupland CAC, Bafadhel M, et al. Development and validation of a new algorithm for improved cardiovascular risk prediction. *Nat Med*. Published online April 18, 2024:1-8. doi:10.1038/s41591-024-02905-y

6. *Cardiovascular Disease: Risk Assessment and Reduction, Including Lipid Modification*. National Institute for Health and Care Excellence; 2023. Accessed June 17, 2024. https://www.nice.org.uk/guidance/ng238

7. *Hypertension in Adults: Diagnosis and Management*. National Institute for Health and Care Excellence; 2019. Accessed June 17, 2024. https://www.nice.org.uk/guidance/ng136

8. Collins GS, Altman DG. Predicting the 10 year risk of cardiovascular disease in the United Kingdom: independent and external validation of an updated version of QRISK2. *BMJ*. 2012;344:e4181. doi:10.1136/bmj.e4181

9. Zhang Z, Rousson V, Lee WC, et al. Decision curve analysis: a technical note. *Annals of Translational Medicine*. 2018;6(15):308-308. doi:10.21037/atm.2018.07.02

10. Talluri R, Shete S. Using the weighted area under the net benefit curve for decision curve analysis. *BMC Medical Informatics and Decision Making*. 2016;16(1):94. doi:10.1186/s12911-016-0336-x

11. Yebyo HG, Aschmann HE, Menges D, Boyd CM, Puhan MA. Net benefit of statins for primary prevention of cardiovascular disease in people 75 years or older: a benefit–harm balance modeling study. *Ther Adv Chronic Dis*. 2019;10:2040622319877745. doi:10.1177/2040622319877745

12. Chalkou K, Vickers AJ, Pellegrini F, Manca A, Salanti G. Decision Curve Analysis for Personalized Treatment Choice between Multiple Options. *Med Decis Making*. 2023;43(3):337-349. doi:10.1177/0272989X221143058

13. Rousson V, Zumbrunn T. Decision curve analysis revisited: overall net benefit, relationships to ROC curve analysis, and application to case-control studies. *BMC Med Inform Decis Mak*. 2011;11:45. doi:10.1186/1472-6947-11-45

14. Assel M, Sjoberg DD, Vickers AJ. The Brier score does not evaluate the clinical utility of diagnostic tests or prediction models. *Diagnostic and Prognostic Research*. 2017;1(1):19. doi:10.1186/s41512-017-0020-3

15. Framingham Heart Study (FHS) | NHLBI, NIH. Accessed April 15, 2024. https://www.nhlbi.nih.gov/science/framingham-heart-study-fhs

16. Van Calster B, McLernon DJ, van Smeden M, et al. Calibration: the Achilles heel of predictive analytics. *BMC Medicine*. 2019;17(1):230. doi:10.1186/s12916-019-1466-7

17. Vickers AJ, Cronin AM, Elkin EB, Gonen M. Extensions to decision curve analysis, a novel method for evaluating diagnostic tests, prediction models and molecular markers. *BMC Med Inform Decis Mak*. 2008;8:53. doi:10.1186/1472-6947-8-53

18. Wynants L, Van Calster B, Collins GS, et al. Prediction models for diagnosis and prognosis of covid-19: systematic review and critical appraisal. *BMJ*. 2020;369:m1328. doi:10.1136/bmj.m1328

19. Markowetz F. All models are wrong and yours are useless: making clinical prediction models impactful for patients. *npj Precis Onc*. 2024;8(1):1-3. doi:10.1038/s41698-024-00553-6

20. Wynants L, Riley RD, Timmerman D, Van Calster B. Random-effects meta-analysis of the clinical utility of tests and prediction models. *Stat Med*. 2018;37(12):2034-2052. doi:10.1002/sim.7653

Figure 1: Change in the value of the weighting function $\omega = \frac{1}{\frac{1}{a-c} + \frac{1}{d-b}}$ over different values of $a - c$ and $d - b$, where the two utilities are in units of quality-adjusted life years added or lost when a patient is flagged positive, depending on whether they are actually positive ($a - c$), or negative ($d - b$).

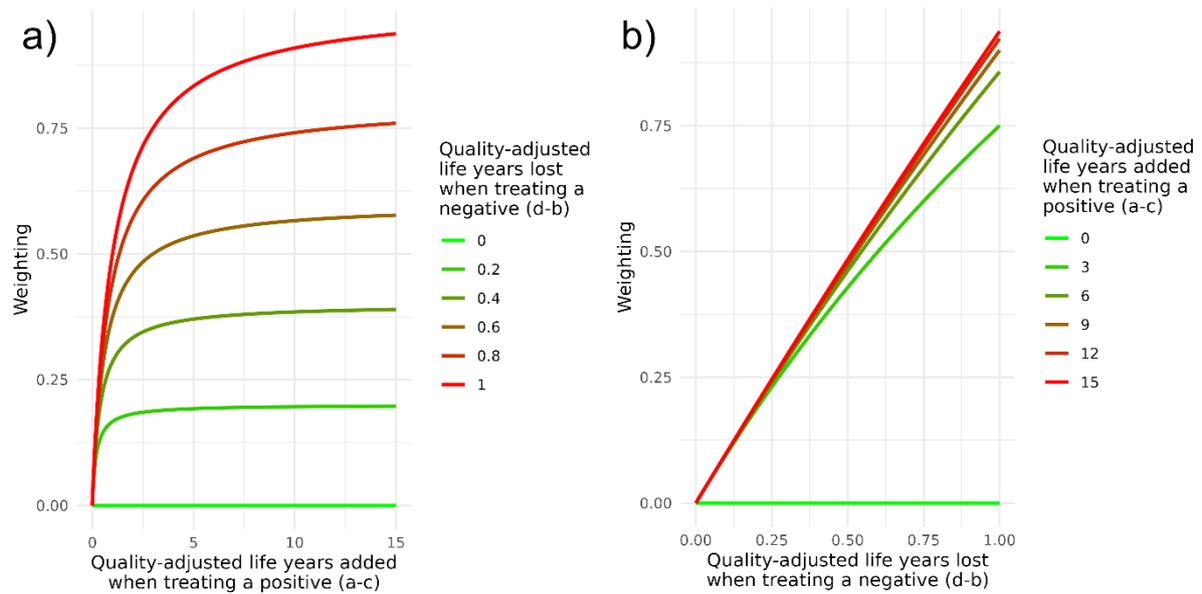

Figure 2: Reported benefit of the cardiovascular model development example when considering multiple treatments. (a) Decision curve of the developed logistic regression model using all (Full.Model) and a limited set (Compact.Model) of the available predictors, as well as the policies of treating all patients (Treat.All) and treating no patients (Treat.No.One). The weighting function to calculate the continuous net benefit function is shown, separated into the components related to the prescription of statins (red) and the recommendation of lifestyle changes (green). (b) Continuous net benefit estimates of all policies for the decisions of prescribing statins and patient management. The unit of the statins component is the benefit of treating a true positive patient with statins per 100 people, and the unit of the overall management component is the benefit of managing a true positive patient as high risk per 100 people.

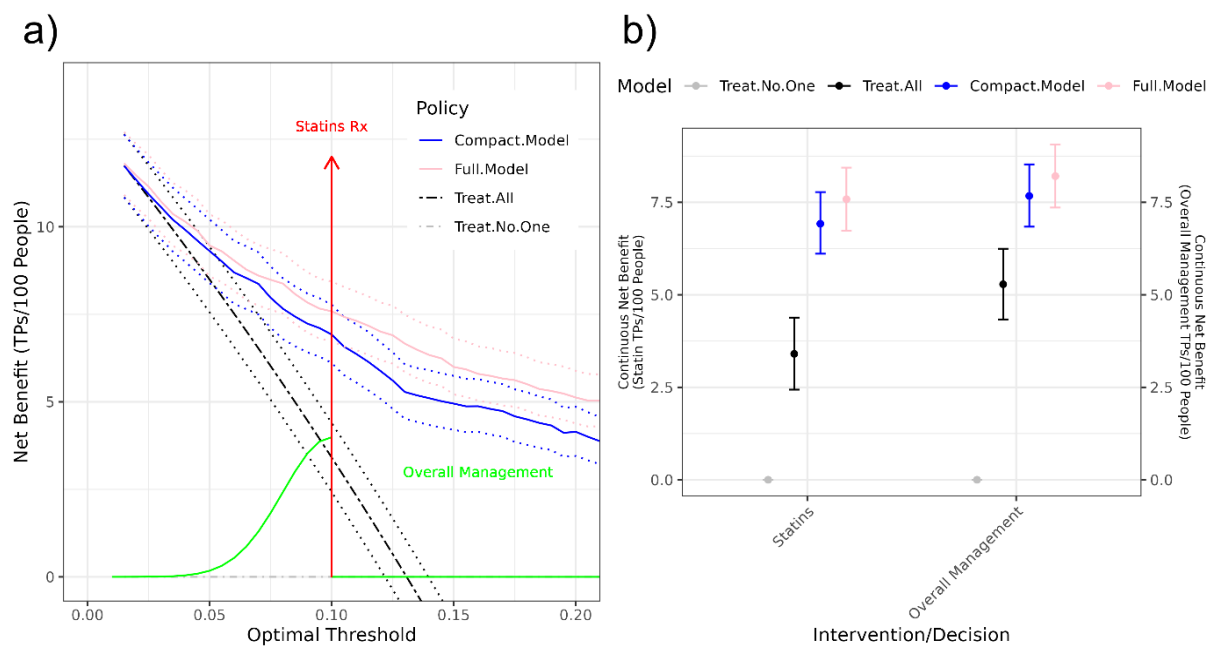

Figure 3: Reported benefit of the cardiovascular model development example when considering a single treatment with a distribution of optimal thresholds across the population. (a) Decision curve of the developed logistic regression model using all (Full.Model) and a limited set (Compact.Model) of the available predictors, as well as the policies of treating all patients (Treat.All) and treating no patients (Treat.No.One). The assumed distribution of optimal thresholds across the population is shown in red. (b) Continuous net benefit estimates of all policies for the decision of prescribing statins, with different assumptions of everyone having an optimal threshold of 10% (left), and individuals having optimal thresholds taken from the distribution shown in (a) (right). The unit is the average benefit of treating a true positive patient with statins per 100 people.

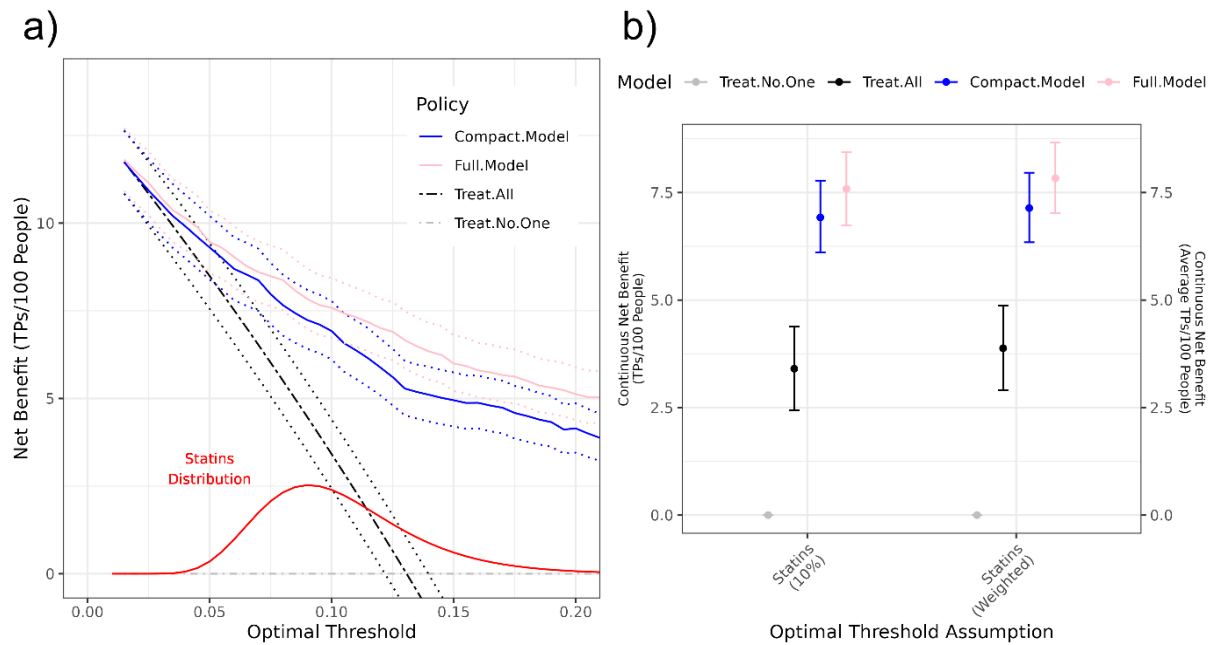

*Supplementary 1.1: Proof that a uniform weighting function yields the likelihood function when calculating the continuous net benefit.*

We seek to prove the statement that, when choosing a uniform weighting function $\omega_{cont}(t') = 1$, the difference in continuous net benefit between two models is equivalent to their difference in the likelihood function. In order to write the proof, we will write the true and false positives for a model $f: X \to Y$ on a dataset $\{x_i, y_i\}_{i=1,2\ldots N}$ as:

$$TP(t) = \sum_i \theta(f(x_i) - t) y_i$$

$$FP(t) = \sum_i \theta(f(x_i) - t)(1 - y_i)$$

Where $\theta$ is a step function equal to 1 if $f(x_i) - t > 0$, and 0 if $f(x_i) - t \leq 0$. We consider two models $f_1, f_2: X \to (0,1)$ with true and false positives $TP_1(t)$, $TP_2(t)$, $FP_1(t)$ and $FP_2(t)$. Their difference in continuous net benefit is thus:

$$NB_{cont1} - NB_{cont2} = \int_0^1 \frac{TP_1(t')}{t'} - \frac{FP_1(t')}{1-t'} dt' - \int_0^1 \frac{TP_2(t')}{t'} - \frac{FP_2(t')}{1-t'} dt'$$

$$= \sum_i \int_0^1 \frac{\theta(f_1(x_i) - t') y_i}{t'} - \frac{\theta(f_1(x_i) - t')(1 - y_i)}{1-t'} dt'$$

$$- \int_0^1 \frac{\theta(f_2(x_i) - t') y_i}{t'} - \frac{\theta(f_2(x_i) - t')(1 - y_i)}{1-t'} dt'$$

$$= \sum_i \int_0^{f_1(x_i)} \frac{y_i}{t'} - \frac{1 - y_i}{1-t'} dt' - \int_0^{f_2(x_i)} \frac{y_i}{t'} - \frac{1 - y_i}{1-t'} dt'$$

$$= \sum_i \int_0^{f_1(x_i)} \frac{y_i}{t'} + \frac{y_i}{1-t'} dt' - \int_0^{f_1(x_i)} \frac{1}{1-t'} dt' - \int_0^{f_2(x_i)} \frac{y_i}{t'} + \frac{y_i}{1-t'} dt'$$

$$+ \int_0^{f_2(x_i)} \frac{1}{1-t'} dt'$$

$$= \sum_i y_i(\log(t') - \log(1 - t') \,|_{t'=0}^{t'=f_1(x_i)} + (\log(1 - t') \,|_{t'=0}^{t'=f_1(x_i)}$$

$$- y_i(\log(t') - \log(1 - t') \,|_{t'=0}^{t'=f_2(x_i)} - (\log(1 - t') \,|_{t'=0}^{t'=f_2(x_i)}$$

At this point, the left extremes of the evaluation of the integrals cancels out as it is equal in the two models. This cancellation can be seen if we instead take the integral between $\epsilon$ and 1, so that $\log(\epsilon)$ is well defined, and take the limit $\epsilon \to 0$ after $\log(\epsilon)$ is cancelled in the difference of net benefit between the two models.

$$NB_{cont1} - NB_{cont2}$$

$$= \sum_i y_i \log(f_1(x_i)) - y_i \log(1 - f_1(x_i)) + \log(1 - f_1(x_i)) - y_i \log(f_2(x_i))$$

$$+ y_i \log(1 - f_2(x_i)) - \log(1 - f_2(x_i))$$

$$= \sum_i y_i \log(f_1(x_i)) + (1 - y_i) \log(1 - f_1(x_i))$$

$$- \sum_i y_i \log(f_2(x_i)) + (1 - y_i) \log(1 - f_2(x_i))$$

$$= \mathcal{L}(f_1, \{x_i, y_i\}_{i=1,2\dots N}) - \mathcal{L}(f_1, \{x_i, y_i\}_{i=1,2\dots N}) \quad \square$$

Where $\mathcal{L}$ is the likelihood function of the model for the observed data.

*Supplementary 1.2: Proof that a parabola weighting function yields the Brier score when calculating the continuous net benefit.*

The proof for this statement starts in the same way as that of Supplementary 1.1. We consider the weighting function $\omega_{cont}(t') = t'(1 - t')$ to calculate the continuous net benefit of a model $f_1$.

$$NB_{cont1} = \int_0^1 t'(1-t')\left(\frac{TP_1(t')}{t'} - \frac{FP_1(t')}{1-t'}\right)dt'$$

$$= \sum_i \int_0^1 \theta(f_1(x_i) - t')y_i(1-t') - \theta(f_1(x_i) - t')(1-y_i)t'dt'$$

$$= \sum_i \int_0^{f_1(x_i)} y_i - y_it' - t' + y_it'dt'$$

$$= \sum_i (y_it' - \frac{1}{2}(t')^2|_{t'=0}^{t'=f_1(x_i)}$$

$$= -\frac{1}{2}\sum_i -2y_if_1(x_i) + f_1(x_i)^2 + y_i^2 - y_i^2$$

$$= -\frac{1}{2}\sum_i \left(y_i - f_1(x_i)\right)^2 + \frac{1}{2}\sum_i y_i^2$$

$$= -\frac{1}{2}Brier\left(f_1, \{x_i, y_i\}_{i=1,2\ldots N}\right) + \frac{1}{2}\pi$$

Where $\pi$ is the prevalence of the outcome in the population. Thus, by comparing the continuous net benefits of two models $f_1$ and $f_2$ we obtain:

$$NB_{cont1} - NB_{cont2} = \frac{1}{2}\left(-Brier\left(f_1, \{x_i, y_i\}_{i=1,2\ldots N}\right) + Brier\left(f_2, \{x_i, y_i\}_{i=1,2\ldots N}\right)\right) \quad \square$$

*Supplementary 1.3: Calculating the expected net benefit for a population with a distribution of utilities a, b, c and d.*

Assume that we have a population $\{x_i, y_i\}_{i=1,2\ldots N} \sim X, Y$ where each individual has a unique set of utilities $\{a_i, b_i, c_i, d_i\}_{i=1,2\ldots N} \sim A, B, C, D$. We assume that we do not know the individual values of $a_i$, $b_i$, $c_i$ and $d_i$ for each individual, nor the distribution of $A$, $B$, $C$ and $D$. Instead, we know the distribution of thresholds $T^*$ where $t^* \sim T^*$ is the optimal threshold for each individual, with $\frac{1-t^*}{t^*} = \frac{a-c}{d-b}$, but do not have individual values $t_i^*$. We consider a prediction model $f : X \to Y$ which gives each individual a score $\{f(x_i)\}_{i=1,2\ldots N}$.

The conditional utility of the model for any arbitrary threshold $t$ is distributed across the population as:

$$U(f(X) = f(x), A = a, B = b, C = c, D = d) = aP(Y = 1, f(x) > t | a, b, c, d)$$
$$+ bP(Y = 0, f(x) > t | a, b, c, d) + cP(Y = 1, f(x) < t | a, b, c, d)$$
$$+ dP(Y = 0, f(x) < t | a, b, c, d)$$

We assume that the coefficients of the utility $A$, $B$, $C$ and $D$ are independent of the performance (true and false positives and negative rates) of the model. This assumption is needed as otherwise the overall proportion of true positives at a particular threshold would not be equal to the proportion of true positives in the subgroup for which that threshold is optimal. This assumption is not always reasonable, as it is likely that patients with a high optimal threshold have different characteristics to those with a low optimal threshold. However, since we do not have the individual thresholds across the population, and instead have a distribution, this assumption needs to be made.

**Key Assumption 1.3.1)** $f(X), Y \perp A, B, C, D$

We use this assumption to remove the conditionals of the utility:

$$U(f(x), a, b, c, d) = aP(y = 1, f(x) > t) + bP(y = 0, f(x) > t)$$
$$+ cP(y = 1, f(x) < t) + dP(y = 0, f(x) < t)$$

We want to approximate the expected estimate across the population.

$$E_{f(x),a,b,c,d \sim X,A,B,C,D} U$$

$$= \int_{f(X),A,B,C,D} p(f(x),a,b,c,d)(aP(y=1,f(x)>t) + bP(y=0,f(x)>t)$$

$$+ cP(y=1,f(x)<t) + dP(y=0,f(x)<t))\mathbf{d}f(x)\mathbf{d}a\mathbf{d}c\mathbf{d}b\mathbf{d}d$$

We have assumed that $f(X) \perp A,B,C,D$, and thus $p(f(x),a,b,c,d) = p(f(x))p(a,b,c,d)$ so:

$$E_{f(x),a,b,c,d \sim f(X),A,B,C,D} U$$

$$= \int_{A,B,C,D} p(a,b,c,d)(a \int_{f(X)} P(y=1,f(x)>t)p(f(x))\mathbf{d}f(x)$$

$$+ b \int_{f(X)} P(y=0,f(x)>t)p(f(x))\mathbf{d}f(x)$$

$$+ c \int_{f(X)} P(y=1,f(x)<t)p(f(x))\mathbf{d}f(x)$$

$$+ d \int_{f(X)} P(y=0,f(x)<t)p(f(x))\mathbf{d}f(x))\mathbf{d}a\mathbf{d}b\mathbf{d}c\mathbf{d}d$$

We can substitute each of the integrals over $f(X)$ over their estimate, i.e., the true and false positives and true and false negatives of the entire population.

$$E_{f(x),a,b,c,d \sim X,A,B,C,D} U = \int_{A,B,C,D} p(a,b,c,d)(aTP(t^*) + bFP(t^*) + cFN(t^*) + dFP(t^*))\mathbf{d}a\mathbf{d}b\mathbf{d}c\mathbf{d}d$$

In order to make the following step, the second key assumption is needed. The integral above is over four degrees of freedom, and could not be solved without knowledge of the distribution $p(a,b,c,d)$, even if we know or assume $p(t^*)$.

Instead, we need to be able to take the expected utility over values of $T^*$, for this to be possible, the values of $a$, $b$, $c$ and $d$ need to be fully determined by $t^*$. In other words, the distribution of the four utilities needs to have a single degree of freedom, which can be considered a latent variable (perhaps, overall health) which needs to have a one-to-one mapping with the optimal threshold $t^* = \frac{1}{1+\frac{a-c}{d-b}}$. In a way, since we can always take a linear transformation of the utility in order to get a meaningful net benefit, we could always set $c$ and $d$ to be constant so that the two utilities have at most two degrees of freedom.

**Key Assumption 1.3.2)** There exists four well-defined functions $g_a$, $g_b$, $g_c$ and $g_d$ so that $a = g_a(t^*)$, $b = g_b(t^*)$, $c = g_c(t^*)$ and $d = g_d(t^*)$.

In this case, we can instead take the expected value over $T^*$, so that we get:

$$E_{f(x),a,b,c,d \sim f(X),A,B,C,D} U = E_{f(x),t^* \sim f(X),T^*} U$$

$$= \int_{T^*} p(t^*)\big(aTP(t^*) + bFP(t^*) + cFN(t^*) + dFP(t^*)\big)\mathbf{d}t^*$$

From this point, the derivation is very similar to that of the continuous net benefit:

$$E_{f(x),t^* \sim f(X),T^*} U = \int_{T^*} \frac{p(t^*)}{\frac{1}{a(t^*) - c(t^*)} + \frac{1}{d(t^*) - b(t^*)}} \left(\frac{1}{t}TP(t^*) - \frac{1}{1-t}FP(t^*)\right)\mathbf{d}t^*$$

In this case, the weighting function $\omega_E(t^*) = \frac{p(t^*)}{\frac{1}{a(t^*)-c(t^*)} + \frac{1}{b(t^*)-d(t^*)}}$ is influenced both by the probability distribution of the optimal thresholds across the population $p(t^*)$, and by the 'importance' of these thresholds, according to the harmonic mean of $a(t^*) - c(t^*)$ and $d(t^*) - b(t^*)$.

*Supplementary 2: Development and validation of the cardiovascular risk prediction models*

In this supplementary, we detail the development and validation of the two cardiovascular models presented in the examples of this study. This supplementary loosely follows the Methods section of the TRIPOD reporting checklist (https://doi.org/10.1186/s12916-014-0241-z). Full code for this analysis is shared in GitHub (https://github.com/jaurioles/continuousNetBenefit).

## S3.1. Sources of data and participants

The development and validation data comprises individuals from the Framingham Heart Study, a prospective observational study in which individuals had baseline predictors recorded and were followed over a 24 years for outcomes related to cardiovascular disease. It is sourced from the riskCommunicator R package (10.32614/CRAN.package.riskCommunicator).

All individuals were considered eligible for the training and internal validation of the model, as long as they had 10 years of follow-up (individuals who died or were otherwise censored before this were excluded from the study). As the Framingham Heart Study ran from 1956 to 1968, no patient in the data was prescribed statins.

## S3.2. Outcome and predictors

The outcome was defined as the development of a new cardiovascular disease event within 10 years after baseline. Two sets of predictors were included, the core predictors used by both models (age at baseline, sex, current smoking status, and prevalent cardiovascular disease), and the added predictors used by the full model (serum total cholesterol, systolic blood pressure, body-mass index, prevalent diabetes, use of anti-hypertensive medication, prevalent hypertension, and casual serum glucose).

*S3.3. Sample size and missing data*

Since the aim of this analysis is not to actually develop and validate a potentially useful prediction model, but to showcase how to calculate and present the net benefit, we did not carry out any sample size calculations. The Framingham Heart Study dataset has been used frequently for the development of clinical prediction models in a didactic environment.

Single imputation, using the mice package in R (10.32614/CRAN.package.mice), was used to impute any missing data, with 30 iterations used.

In order to account for the possibility of dependent censoring biasing the model's predictions and the estimation of the performance metrics, inverse probability of censoring weighting (IPCW, https://doi.org/10.1177/0962280216628900), was used in order to up-weight individuals more likely to have been censored by 10 years of follow-up. This weighting was applied both when fitting the models and when calculating all the presented performance metrics. The issue of competing risks was not explored in this example. The full set of predictors used by the models was also used to build the propensity score model.

*S3.4. Statistical analysis methods*

Two logistic regression models were developed, one using only the core predictors, and one using all predictors. In order to account for the potential non-linear effect of age, a penalised smoothed spline with three degrees of freedom was used for age.

The continuous net benefit was calculated as outlined in the main manuscript with the different weighting functions outlined in the examples. The optimism and 95% confidence intervals of the net benefit, continuous net benefit and decision curves were calculated in the same way. Optimism was corrected through 5000 bootstraps, which were used to calculate the difference between the apparent and actual performance of the models. 95% confidence intervals were also calculated through 5000 bootstraps.